\documentclass[12pt]{article}

\usepackage{authblk}
\usepackage{graphicx, setspace, amsmath, amsfonts, amssymb, fontenc, secdot, rotating,gensymb, makecell}

\usepackage[top=1in, bottom=1in, left=0.7in, right=0.7in]{geometry}
\usepackage[utf8]{inputenc}
\usepackage{pdflscape}
\def\bp{\begin{pmatrix}}
\def\ep{\end{pmatrix}}
\def\bc{\begin{center}}
\def\ec{\end{center}}
\def\be{\begin{equation}}
\def\ee{\end{equation}}

\begin{document}
\title{Probing 4 $\times$ 4 quark mixing matrix }

\author{Gurjit Kaur, Gulsheen Ahuja$^*$, Dheeraj Shukla and Manmohan Gupta$^\dagger$\\Department of Physics,\\ Panjab University, Chandigarh.\\
\vspace{0.7cm}
* gulsheen@pu.ac.in\\
$^\dagger$ mmgupta@pu.ac.in}

\onehalfspacing
\maketitle

\begin{abstract}
Without adhering to any specific model, we have presented 4 $\times$ 4 quark mixing matrix as an extension of the 3 $\times$ 3 PDG parametrization of the CKM matrix. Using unitarity constraints as well as the hierarchy among the elements of the 3 $\times$ 3 CKM matrix, we have found the hierarchy among the 4$^{th}$ row and 4$^{th}$ column elements of the 4 $\times$ 4 quark mixing matrix. Further, for the fourth generation case, we have explicitly found the 9 independent rephasing invariant parameters $J_{4\times 4}$. Also, using phenomenological estimates of the 4$^{th}$ row and 4$^{th}$ column elements, we have numerically evaluated these 9 parameters.  
\end{abstract}

\section{Introduction}
Over the last few decades, quark mixing phenomenology, encoded in Cabibbo-Kobayashi-Maskawa  (CKM) \cite{ckm1,ckm2} paradigm has registered remarkable progress. On the theoretical front, CKM paradigm along with the unitarity of the CKM matrix has played a crucial role in understanding several important features of flavor physics. On the experimental front,  several groups like Particle Data Group (PDG) \cite{pdg24}, CKMfitter \cite{ckmfit}, HFLAV \cite{hflav23}, UTfit \cite{utfit}, etc., have been actively engaged in continuously updating their analyses to arrive at more and more refined conclusions regarding CKM parameters. 

Inspite of remarkable progress in the context of CKM phenomenology, at present, we are saddled with several  issues which need to be addressed. 
Recently, PDG \cite{pdg24} has reported a $2.3\sigma$ deviation from unitarity in the first row of the mixing matrix, i.e.,
\be 
|V_{ud}|^2+|V_{us}|^2+|V_{ub}|^2=0.9984\pm0.0007 .\ee 
This gives rise to  two intriguing anomalies related to the element $|V_{us}|$ or the Cabibbo mixing angle, known as the Cabibbo Angle Anomaly 1 (CAA1) and the Cabibbo Angle Anomaly 2 (CAA2)  \cite{2302.14097}. The former identifies the above mentioned tension regarding first row of the mixing matrix, while, the latter refers to a $3.1 \sigma$ discrepancy in two different measurements of $V_{us}$, e.g., $0.22308(55)$ and $0.22536(47)$ from semi-leptonic $Kl_3$ decay and from the ratio of kaon and pion leptonic decay rates respectively \cite{flav21}.

The violation noted in the unitarity of the first row of the CKM matrix elements, if confirmed, would be a clear indication of New physics \cite{caa1np1, caa1np2}. Perhaps, the simplest way to resolve this anomaly is to extend the Standard Model (SM) to the fourth generation of fermions. The extension of SM by introducing additional family of quarks has been investigated before \cite{brancovec}-\cite{4genburas}. Various phenomenological analyses \cite{4genpheno1}-\cite{4genpheno3} have been carried out the with the inclusion of the fourth generation to the SM.
It should also be noted that even in the leptonic sector, in order to accommodate additional neutrinos, an extension to the fourth generation has been considered \cite{xingfourlep1, xingfourlep2}. 

In the present work, without adhering to any specific model, considering the 4 $\times$ 4 generalization of the 3 generation CKM matrix, we have studied the implications of unitarity. Specifically, we have presented the 4 $\times$ 4 quark mixing matrix as an extension of the 3 $\times$ 3 PDG parametrization of the CKM matrix which recently has been shown to be a preferred one for carrying out phenomenological analyses \cite{ourptep}. Interestingly, the unitarity implications of hierarchy of the 3 $\times$ 3 CKM matrix elements have also been recently investigated \cite{ourhier}. In the present work, we have made an attempt to carry out a similar exercise for the 4 $\times$ 4 quark mixing matrix. Further, we have also formulated the rephasing invariant parameter in the fourth generation, referred to as $J_{4\times 4}$. Using the various constraints available in the literature for the 4 $\times$ 4 quark mixing matrix, we have carried out numerical evaluation of $J_{4\times 4}$.

 \section{4$\times$4 quark mixing matrix in terms of mixing angles and phases }
In the present work, in order to extend the 3$\times$3 quark mixing matrix to the fourth generation, we use the commonly known primed notation, i.e. $t^\prime$ and $b^\prime$ for fourth generation quarks.
The SM with an additional fourth generation is one of the simplest extensions of the SM and retains all of its essential
features, i.e., it obeys all the SM symmetries and does not introduce any new ones. This addition of the fourth generation to the SM leads to an extension of the 3 $\times$ 3 Cabibbo-Kobayashi-Maskawa (CKM) matrix, given by
\be 
\bp
V_{ud}& V_{us}& V_{ub}\\
V_{cd}& V_{cs}& V_{cb}\\
V_{td}& V_{ts}& V_{tb}
\ep \ee
to a 4 $\times$ 4 quark mixing matrix, referred to as CKM4, expressed as 
\be \begin{pmatrix}
V_{ud}&V_{us}&V_{ub}&V_{ub^\prime}\\
V_{cd}&V_{cs}&V_{cb}&V_{cb^\prime}\\
V_{td}&V_{ts}&V_{tb}&V_{tb^\prime}\\
V_{t^\prime d}&V_{t^\prime s}&V_{t^\prime b}&V_{t^\prime b^\prime}\\
\end{pmatrix}.\label{ckm4}\ee

This 4 $\times$ 4 quark mixing matrix can be parametrized in terms of 6 real parameters and 3 phases. To this end, we consider an n-dimensional unitary matrix $U_{n\times n}$, expressed as \cite{genuni} 
 \be  U_{n\times n}=[U_{12}(\theta_{12},\sigma_{12})~.~U_{13}(\theta_{13},\sigma_{13})............U_{1(n-1)}(\theta_{1(n-1)},\sigma_{1(n-1)})~.~U_{1n}(\theta_{1n},\sigma_{1n})]~.~ [U^\prime],\label{eq.nge}\ee
where $~U_{12}(\theta_{12},\sigma_{12})~,~U_{13}(\theta_{13},\sigma_{13})............U_{1(n-1)}(\theta_{1(n-1)},\sigma_{1(n-1)})~,~U_{1n}(\theta_{1n},\sigma_{1n})~$ represent rotation matrices which may be denoted as $U_{pq}(\theta_{pq},\sigma_{pq})$, with  $p$ and $q$ being any two numbers with $q > p$.
 These $U_{pq}$'s are n-dimensional unimodular unitary matrices with 
 the diagonal elements being 1 except the $p^{th}\times p^{th} $ and  $q^{th}\times q^{th} $ elements, which are $c_{pq} = \cos \theta_{pq}$.
Further, for these unimodular matrices, all the off - diagonal elements are zero except the $p^{th}\times q^{th}$ element which is $-s_{pq} e^{-\iota \sigma_{pq}}$  and the $q^{th}\times p^{th}$ element which is $s_{pq} e^{\iota \sigma_{pq}}$, where $s_{pq}=\sin\theta_{pq}$.

For example, for n=3, $U_{12}(\theta_{12},\sigma_{12})$ is
\be U_{12}=\bp 
c_{12}  & -s_{12} e^{-\iota \sigma_{12}}&0\\
s_{12} e^{\iota \sigma_{12}} &c_{12}&0\\
0&0&1
\ep .
 \ee
Further, in equation (\ref{eq.nge}), $U^\prime$ is given by
\be
U^\prime=\bp
e^{\iota \delta_n}&0\\0&U_{(n-1)\times (n-1)}
\ep .\ee
For n=3, it becomes
\be U^\prime=\bp
e^{\iota \delta_3}&0\\0&U_{2\times 2}
\ep ~~~~~
\text{with}~~~~~
 U_{2\times 2}  =\bp e^{\iota \delta_1}&0\\
0&e^{\iota \delta_2}
\ep 
\bp
c_\theta  & -s_\theta e^{-\iota \sigma}\\
s_\theta e^{\iota \sigma} &c_\theta 
 \ep, \ee
leading to \be U^\prime=\bp e^{\iota \delta_1}&0&0\\
0&e^{\iota \delta_2}&0\\
0&0&e^{\iota \delta_3}
\ep \bp 
1&0&0\\
0&c_\theta  & -s_\theta e^{-\iota \sigma}\\
0&s_\theta e^{\iota \sigma} &c_\theta 
\ep,
\ee
expressed as
  \be
 U^\prime~~=~~D~(\delta_1,\delta_2, \delta_3)~.~U_{23}(\theta,\sigma)
\ee
Therefore, for n=3, equation (\ref{eq.nge}) can be written as
\be U_{3\times 3}=[ U_{12}(\theta_{12},\sigma_{12})~.~U_{13}(\theta_{13},\sigma_{13})]~.~[D(\delta_1,\delta_2, \delta_3)~.~U_{23}(\theta_{23},\sigma_{23})],\ee
re-written as \be U_{3\times 3}=D(\delta_1,\delta_2, \delta_3)~.~ U_{12}(\theta_{12},\sigma_{12})~.~U_{13}(\theta_{13},\sigma_{13})~.~U_{23}(\theta_{23},\sigma_{23}).\ee
Therefore, a 3$\times$3 unitary matrix can be expressed as a product of 3 rotation matrices and a diagonal phase matrix.

Similarly, for the case of n=4, equation (\ref{eq.nge}) is given by
\be  U_{4\times 4}=[ U_{12}(\theta_{12},\sigma_{12})~.~U_{13}(\theta_{13},\sigma_{13})~.~U_{14}(\theta_{14},\sigma_{14})]~.~[D(\delta_1,\delta_2, \delta_3,\delta_4)
~.~ U_{23}(\theta_{23},\sigma_{23})~.~U_{24}(\theta_{24},\sigma_{24})~.~U_{34}(\theta_{34},\sigma_{34})],
\ee
expressed as \be  U_{4\times 4}= D(\delta_1,\delta_2, \delta_3,\delta_4)
~.~U_{12}(\theta_{12},\sigma_{12})~.~U_{13}(\theta_{13},\sigma_{13})~.~U_{14}(\theta_{14},\sigma_{14})~.~U_{23}(\theta_{23},\sigma_{23})~.~U_{24}(\theta_{24},\sigma_{24})~.~U_{34}(\theta_{34},\sigma_{34}) .
\ee
The above represents a 4$\times$4 unitary matrix expressed as a product of 6 rotation matrices and a diagonal phase matrix.

One can arrive at several representations of 4$\times$4 quark mixing matrix, in the present work, we present a particular one which can be easily reduced to the PDG representation of the 3$\times$3 quark mixing matrix. 
This representation of the 4$\times$4 quark mixing matrix will be referred to as PDG4. It may be mentioned that in order to obtain the PDG representation, one needs to take product of the 3 rotation matrices $U_{23},~U_{13},~U_{12}$, multiplied in this particular sequence. Now, in order to obtain PDG4, we need to take product of 6 rotation matrices, i.e., along with $U_{23},~U_{13},~U_{12}$, we also consider $U_{34},~U_{24},~U_{14}$. 

It may be noted that in the case of PDG representation, the three mixing angles correspond to three experimentally known CKM matrix elements, i.e., \be s_{12} \cong V_{us}, ~~ s_{13} \cong V_{ub},~~s_{23} \cong V_{cb}, \ee therefore, likewise, in the case of the fourth generation matrix also it is desirable that one gets \be s_{14} \cong V_{ub^\prime},~~s_{24} \cong V_{cb^\prime},~~s_{34} \cong V_{tb^\prime}.\label{approx} \ee 
Therefore, the above mentioned 6 rotation matrices need to be multiplied in a particular sequence
\be V_{PDG_4}= D(\delta_1,\delta_2,\delta_3,\delta_4).~U_{34}(\theta_{34},\sigma_{34}).~U_{23}(\theta_{23},\sigma_{23}).~U_{24}(\theta_{24},\sigma_{24}).~U_{13}(\theta_{13},\sigma_{13}).~U_{12}(\theta_{12},\sigma_{12}).~U_{14}(\theta_{14},\sigma_{14}).\ee
The above 4$\times$4 mixing matrix involves 6 mixing angles $\theta_{12},~\theta_{13},~\theta_{23},~\theta_{14},~\theta_{24}~\text{and}~ \theta_{34}$ as well as 10 phases $\delta_1,\delta_2,\delta_3,\delta_4$ and $\sigma_{12},~\sigma_{13},~\sigma_{23},~\sigma_{14},~\sigma_{24}$, $\sigma_{34}$. Out of these 10 phases, 7 can be removed using the facility of rephasing. The remaining 3 phases may be defined as
 $\sigma_{ub}=(\text{$\sigma $}_{12}-\text{$\sigma $}_{13}+\text{$\sigma $}_{23}),~\sigma_{cb^\prime}=(\text{$\sigma $}_{24}-\text{$\sigma $}_{23}-\text{$\sigma $}_{34})~ \text{and}~ \sigma_{ub^\prime}= (\text{$\sigma $}_{14}-\text{$\sigma $}_{12}-\text{$\sigma $}_{23}-\text{$\sigma $}_{34})$.

Therefore, one gets \begin{small}
$$V_{PDG_4}= \left(
\begin{array}{cccc}
\vspace{2cm}
 c_{12} c_{13} c_{14} 
 & -c_{13} c_{14} s_{12}
  & -e^{-i \text{$\sigma $ub}} c_{14} s_{13} 
  & -e^{-i \text{$\sigma ub^\prime$}} s_{14} \\ 
  
 c_{23} c_{24} s_{12}                & c_{12} c_{23} c_{24} & e^{-i \text{$\sigma $ub}+i \text{$\sigma ub^\prime$}-i \text{$\sigma cb^\prime$}} s_{13} s_{14} s_{24}                 & -e^{-i \text{$\sigma cb^\prime$}} c_{14} s_{24} \\ 
 
 -e^{i \text{$\sigma $ub}} c_{12} c_{24} s_{13} s_{23}   & +e^{i \text{$\sigma $ub}} s_{12} s_{13} s_{23} c_{24}           & -c_{13} c_{24} s_{23} & \\
 
 \vspace{2cm}
 -e^{i \text{$\sigma ub^\prime$}-i \text{$\sigma cb^\prime$}} c_{12} c_{13} s_{14} s_{24}          &+e^{i \text{$\sigma ub^\prime$}-i \text{$\sigma cb^\prime$}} c_{13} s_{12} s_{14} s_{24}            & & \\

 e^{i \text{$\sigma $ub}} c_{12} c_{23} c_{34} s_{13}      & -e^{i \text{$\sigma $ub}} c_{23} c_{34} s_{12} s_{13}                     & c_{13} c_{23} c_{34} & -c_{14} c_{24} s_{34}\\   
 
+e^{i \text{$\sigma $ub}+i \text{$\sigma cb^\prime$}} c_{12} s_{23} s_{24} s_{34} s_{13}       
              &-e^{i \text{$\sigma $ub}+i \text{$\sigma cb^\prime$}} s_{12} s_{23} s_{24} s_{34} s_{13}   &+e^{i \text{$\sigma ub^\prime$}-i \text{$\sigma $ub}} c_{24} s_{13} s_{14} s_{34}                      &   \\ 

   +c_{34} s_{12} s_{23}   &        +c_{12} c_{34} s_{23} &
 +e^{i \text{$\sigma cb^\prime$}} c_{13} s_{23} s_{24} s_{34}                 & \\
 
 -e^{i \text{$\sigma ub^\prime$}} c_{12} c_{13} c_{24} s_{14} s_{34}
 &+e^{i \text{$\sigma ub^\prime$}} c_{13} c_{24} s_{12} s_{14} s_{34} & &\\  \vspace{2cm}
 
 -e^{i \text{$\sigma cb^\prime$}} c_{23} s_{12} s_{24} s_{34} &-e^{i \text{$\sigma cb^\prime$}} c_{12} c_{23} s_{24} s_{34}         & &\\
  
 e^{i \text{$\sigma ub^\prime$}} c_{12} c_{13} c_{24} c_{34} s_{14} & -e^{i \text{$\sigma ub^\prime$}} c_{13} c_{24} c_{34} s_{12} s_{14} & -e^{i \text{$\sigma ub^\prime$}-i \text{$\sigma $ub}} c_{24} c_{34} s_{13} s_{14}           & c_{14} c_{24} c_{34}\\
 
+e^{i \text{$\sigma cb^\prime$}} c_{23} c_{34} s_{12} s_{24}             & +e^{i \text{$\sigma cb^\prime$}} c_{12} c_{23} c_{34} s_{24}                                            &-e^{i \text{$\sigma cb^\prime$}} c_{13} c_{34} s_{23} s_{24}   &\\
 
 -e^{i \text{$\sigma $ub}+i \text{$\sigma cb^\prime$}} c_{12} c_{34} s_{13} s_{23} s_{24} &                
 +e^{i \text{$\sigma $ub}+i \text{$\sigma cb^\prime$}} c_{34} s_{12} s_{13} s_{23} s_{24}
&
 +c_{13} c_{23} s_{34} 
 &
  \\
  
  +e^{i \text{$\sigma $ub}} c_{12} c_{23} s_{13} s_{34}& -e^{i \text{$\sigma $ub}} c_{23} s_{12} s_{13} s_{34} & &\\
  
  +s_{12} s_{23} s_{34} & +c_{12} s_{23} s_{34}& &\\
\end{array}
\right).$$ \end{small}
It can be easily checked that by ignoring the contribution of the $4^{th}$ generation, the above matrix gets reduced to PDG representation of the 3$\times$3 CKM matrix, expressed in terms of sines and cosines of the three mixing angles  $\theta_{12},~\theta_{13},~ \text{and}~\theta_{23}$ and one phase $\sigma_{ub}$, usually mentioned as $\delta$.

\section{Hierarchical order of CKM4 matrix elements}
Very recently, incorporating unitarity constraints, the hierarchy amongst the 3$\times$3 CKM matrix elements has been rigorously explored using the PDG parameterization \cite{ourhier}. 
In the absence of any definite clues about the $4^{th}$ generation elements of CKM4, to obtain the hierarchical constraints on these, we begin with the hierarchy of 3 generation CKM matrix as observed experimentally \cite{pdg24}, given by
 \be |V_{tb}|> |V_{ud}|>|V_{cs}| >|V_{us}| > |V_{cd}|>|V_{cb}|>|V_{ts}|> |V_{td}|>|V_{ub}| \label{all9}. \ee

 \begin{table}[h!]
\renewcommand{\arraystretch}{1.5}
\centering
\caption{Magnitudes of the 3$\times$3 CKM mixing elements given by PDG (2024)}
\vspace{0.5cm}

\begin{tabular}{|c|c|c|}
\hline
S.No. & Matrix element & Experimental value \cite{pdg24}\\ \hline
1& $V_{ud}$ & $0.97367\pm0.00032$\\ \hline
2& $V_{us}$ & $0.22431\pm0.00085$\\ \hline
3& $V_{ub}$ & $(3.82\pm0.20)\times 10^{-3}$\\ \hline
4& $V_{cd}$ & $0.221\pm0.004$\\ \hline
5& $V_{cs}$ & $0.975\pm0.006$\\ \hline
6& $V_{cb}$ & $0.0411\pm0.0012$\\ \hline
7& $V_{td}$ & $0.0086\pm0.0002$\\ \hline
8& $V_{ts}$ & $0.0415\pm0.0009$\\ \hline
9& $V_{tb}$ & $1.010\pm0.027$\\ \hline

\end{tabular} 
\label{pdginputs}
\end{table}

For the 4$\times$4 CKM matrix as defined in equation (\ref{ckm4}), from unitarity of the first row and first columns of this matrix, we have
 \be |V_{ud}|^2+|V_{us}|^2+|V_{ub}|^2+|V_{ub^\prime}|^2 = 1 
 ,\label{r1}\ee
  \be |V_{ud}|^2+|V_{cd}|^2+|V_{td}|^2+|V_{t^\prime d}|^2 = 1 
 ,\label{c1}\ee
 Subtracting these equations, we get
 \be |V_{t^\prime d}|^2-|V_{u b^\prime}|^2=(|V_{us}|^2-|V_{cd}|^2)+(|V_{ub}|^2-|V_{td}|^2). \label{tpdubp}\ee
Using other combinations of  normalization relations of CKM4, we
get
 \begin{align} |V_{cb^\prime }|^2-|V_{t^\prime s}|^2 & =(|V_{us}|^2-|V_{cd}|^2)+(|V_{ts}|^2-|V_{cb}|^2)~,\label{cbptps}\\
 |V_{t^\prime b}|^2-|V_{tb^\prime }|^2&=(|V_{td}|^2-|V_{ub}|^2)+(|V_{ts}|^2-|V_{cb}|^2) ~,\label{tpbtbp}\\
  |V_{ub^\prime }|^2-|V_{t^\prime s}|^2&=(|V_{cs}|^2+|V_{ts}|^2)-(|V_{ud}|^2+|V_{ub}|^2) ~,\label{ubptps} \\
   |V_{ub^\prime }|^2-|V_{t^\prime b}|^2 &=(|V_{cb}|^2+|V_{tb}|^2)-(|V_{ud}|^2+|V_{us}|^2) ~,\label{ubptpb}\\
  |V_{t^\prime d}|^2 -|V_{cb^\prime }|^2&=(|V_{cs}|^2+|V_{cb}|^2)-(|V_{ud}|^2+|V_{td}|^2) ~,\label{tpdcbp}\\
    |V_{cb^\prime }|^2-|V_{t^\prime b}|^2 &=(|V_{ub}|^2+|V_{tb}|^2)-(|V_{cd}|^2+|V_{cs}|^2)~,\label{cbptpb} \\
  |V_{t^\prime d}|^2 -|V_{tb^\prime }|^2 &=(|V_{ts}|^2+|V_{tb}|^2)-(|V_{ud}|^2+|V_{cd}|^2)~,\label{tpdtbp}\\
  |V_{t^\prime s}|^2 -|V_{tb^\prime }|^2 &=(|V_{td}|^2+|V_{tb}|^2)-(|V_{us}|^2+|V_{cs}|^2)~,\label{tpstbp} \\
|V_{t^\prime s}|^2 -|V_{t^\prime b }|^2 &= (|V_{ub}|^2+|V_{cb}|^2+|V_{tb}|^2)-(|V_{us}|^2+|V_{cs}|^2+|V_{ts}|^2)~,\label{tpstpb}\\
 |V_{ub^\prime }|^2 -|V_{cb^\prime }|^2&=(|V_{cd}|^2+|V_{cs}|^2+|V_{cb}|^2)-(|V_{ud}|^2+|V_{us}|^2+|V_{ub}|^2)~.\label{ubpcbp} \end{align}
  
Using the experimentally determined magnitudes of the CKM mixing elements, presented in Table \ref{pdginputs}, one can obtain useful hints regarding hierarchical pattern among the newly introduced fourth generation elements. 
For example, in equation (\ref{tpdubp}), using the experimental values of $|V_{us}|,|V_{cd}|,|V_{ub}| ~\text{and}~|V_{td}|$, one gets
\be|V_{t^\prime d}|-|V_{ub^\prime } |= 0.001415, ~\text{implying}~
|V_{t^\prime d}|>|V_{ub^\prime } |.\ee
Similarly, using equations (\ref{cbptps})-(\ref{ubpcbp}), we obtain 
\be |V_{c b^\prime }|-|V_{t^\prime s}|=0.001507, ~\text{implying}~ |V_{c b^\prime }|>|V_{t^\prime s}|,\ee
 \be | V_{t^\prime b}|-|V_{tb^\prime}|=0.000092,~\text{implying}~| V_{t^\prime b}|>|V_{tb^\prime}|, \ee
  \be |V_{ub^\prime}|-|V_{t^\prime s}|=0.004299 ,~\text{implying}~ |V_{ub^\prime}|>|V_{t^\prime s}|, \ee
  \be |V_{ub^\prime}|-|V_{t^\prime b}|=0.023441 ,~\text{implying}~ |V_{ub^\prime}|>|V_{t^\prime b}|, \ee
  \be |V_{t^\prime d}|-|V_{c b^\prime }|= 0.004207 ,~\text{implying}~ |V_{t^\prime d}|>|V_{c b^\prime }|, \ee
  \be  |V_{c b^\prime}| -|V_{t^\prime b}|=0.020649, ~\text{implying}~ |V_{c b^\prime}| >|V_{t^\prime b}|, \ee
  \be |V_{t^\prime d}|-|V_{t b^\prime }|=0.024948, ~\text{implying}~ |V_{t^\prime d}|>|V_{t b^\prime }|, \ee
  \be | V_{t^\prime s}|-|V_{tb^\prime}|=0.019234  , ~\text{implying}~| V_{t^\prime s}|>|V_{tb^\prime}|, \ee
 \be |V_{t^\prime s}|-|V_{t^\prime b} |= 0.019142 , ~\text{implying}~ |V_{t^\prime s}|>|V_{t^\prime b} |,\ee
 \be | V_{ub^\prime }|-|V_{cb^\prime }|=0.002792  ,~\text{implying}~ | V_{ub^\prime }|>|V_{cb^\prime }|. \ee
Using the above inequalities, one gets the following hierarchy amongst the newly introduced fourth generation elements
\be V_{t^\prime d}>V_{ub^\prime }>V_{cb^\prime}>V_{t^\prime s}>V_{t^\prime b}>V_{t b^\prime }.\ee
Understanding that the diagonal elements of the 3 generation CKM matrix are much larger than the off diagonal ones, we assume the matrix element $V_{t^\prime b^\prime}$ of the CKM4 is again larger than the other 4$^{th}$ row and 4$^{th}$ column elements, implying
\be V_{t^\prime b^\prime}>V_{t^\prime d}>V_{ub^\prime }>V_{cb^\prime}>V_{t^\prime s}>V_{t^\prime b}>V_{t b^\prime }.\ee
This hierarchy, obtained using the experimentally known CKM matrix elements as well as the unitarity of CKM4 can, thereby, provide vital clues for the measurement of the CKM4 elements.

\section{Rephasing invariant parameters $J_{4\times 4}$ and quartets}
In the earlier sections, we have formulated the PDG4 representation of the $4\times 4$ quark mixing matrix as well as have proposed hierarchical relation among the fourth generation mixing matrix elements. Before finding explicit expressions for rephasing invariants in the 4 generation case, similar to the Jarlskog's rephasing invariant parameter J in the 3 generation case, we first discuss, in detail, the 3 generation case. From this, one can then arrive at the rephasing invariant parameters in the fourth generation, denoted here as $J_{4\times 4}$.
For the 3 generation case, the 6 orthogonality conditions of the CKM matrix are given by
\be{\displaystyle \sum _{\alpha}V_{i\alpha}V_{j\alpha}^{*}=\sum _{i}V_{i\alpha}V_{i\beta}^{*}=0,}\label{ortho} \ee 
where the Greek subscripts run over (d, s, b ) and the Latin superscripts run over (u, c, t).
Using the unitarity between the first two rows of the CKM matrix, we get
 \be V_{ud}V_{cd}^*+ V_{us}V_{cs}^*+ V_{ub}V_{cb}^*=0. \label{uni12}\ee
 Multiplying this equation by $ V_{ud}^*V_{cd}$, one obtains
  \be |V_{ud}|^2|V_{cd}|^2+ V_{us}V_{cs}^* V_{ud}^*V_{cd}+ V_{ub}V_{cb}^* V_{ud}^*V_{cd}=0,\ee
Imaginary part of this equation is
\be Im[V_{us}V_{cs}^* V_{ud}^*V_{cd}]+Im[ V_{ub}V_{cb}^* V_{ud}^*V_{cd}]=0, \ee
Equation (\ref{uni12}) can also be multiplied by $ V_{us}^*V_{cs}$ and by $ V_{ub}^*V_{cb}$, giving 
\be Im[V_{ud}V_{cd}^* V_{us}^*V_{cs}]+Im[ V_{ub}V_{cb}^* V_{us}^*V_{cs}]=0 ~\text{and} \ee
\be Im[V_{ud}V_{cd}^* V_{ub}^*V_{cb}]+Im[ V_{us}V_{cs}^* V_{ub}^*V_{cb}]=0 ~\text{respectively}. \ee
Similarly, using the other 5 orthogonal conditions, one gets 15  more such relations, making a total of 18. From these 18 relations, one arrives at 9 different quartets, i.e., 
\be \begin{split} 
Im[V_{ud}V_{cd}^* V_{us}^*V_{cs}],~~~
Im[ V_{us}V_{cs}^* V_{ub}^*V_{cb}],~~~
Im[V_{ud}V_{cd}^* V_{ub}^*V_{cb}],\\
Im[ V_{cd}V_{td}^*V_{cs}^*V_{ts}],~~~
Im[ V_{cs}V_{ts}^* V_{cb}^*V_{ub}],~~~
Im[V_{cd}V_{td}^* V_{cb}^*V_{tb}],\\
Im[V_{ud}V_{td}^* V_{us}^*V_{ts}],~~~
Im[V_{us}V_{ts}^* V_{ub}^*V_{tb}],~~~
Im[V_{ud}V_{td}^* V_{ub}^*V_{tb}],\end{split}
\label{3gj} \ee
It can be easily checked that these above mentioned 9 quartets are all equal and may be denoted as $  |Im(V_{i\alpha}V_{j\beta}V^*_{i\beta}V^*_{j\alpha})|$, where the Greek subscripts run over (d, s, b ) and the Latin superscripts run over (u, c, t). These are explicitly parametrization independent and denote the CP violating Jarlskog's rephasing invariant parameter J, given by
\be J= |Im(V_{i\alpha}V_{j\beta}V^*_{i\beta}V^*_{j\alpha})|. \label{j}\ee 
 This parameter can also be defined in terms of mixing angles and the CP violating phase. For example, for the PDG representation Jarlskog rephasing invariant can be written as $$J= \sin\theta_{12}\sin \theta_{23}\sin\theta_{13}\cos\theta_{12}\cos\theta_{23}\cos^2\theta_{13} \sin \delta.$$

Coming to the case of fourth generation, following the same procedure as followed for the 3 generation case, we start from the 12 orthogonality conditions of the 4$\times$4 matrix given by
\be{\displaystyle \sum _{\alpha}V_{i\alpha}V_{j\alpha}^{*}=\sum _{i}V_{i\alpha}V_{i\beta}^{*}=0,}\label{ortho4} \ee 
where the Greek subscripts now run over (d, s, b, b$^{\prime}$) and the Latin superscripts run over (u, c, t, t$^{\prime}$). Using the orthogonality between the first two rows of the mixing matrix, we get
\be V_{ud}V_{cd}^*+ V_{us}V_{cs}^*+ V_{ub}V_{cb}^*+V_{ub^\prime}V_{cb^\prime}=0.\label{4uni12}\ee 
Multiplying the above equation by 
$ V_{ud}^*V_{cd}$, one gets 
\be Im[ V_{us}V_{cs}^*V_{ud}^*V_{cd}]+Im[ V_{ub}V_{cb}^*V_{ud}^*V_{cd}]+Im[V_{ub^\prime}V_{cb^\prime}V_{ud}^*V_{cd}]=0.\label{4ri1}\ee
Equation (\ref{4uni12}) may also be multiplied by $ V_{us}^*V_{cs}$, by $ V_{ub}^*V_{cb}$ and by $V_{ub^\prime}V_{cb^\prime}$, giving 
\be Im[ V_{ud}V_{cd}^*V_{us}^*V_{cs}]+Im[ V_{ub}V_{cb}^*V_{us}^*V_{cs}]+Im[V_{ub^\prime}V_{cb^\prime}V_{us}^*V_{cs}]=0,\label{ri2}\ee
\be Im[ V_{ud}V_{cd}^*V_{ub}^*V_{cb}]+Im[ V_{us}V_{cs}^*V_{ub}^*V_{cb}]+Im[V_{ub^\prime}V_{cb^\prime}V_{ub}^*V_{cb}]=0~\text{and}\label{ri3}\ee
\be Im[ V_{ud}V_{cd}^*V_{ub^\prime}^*V_{cb^\prime}]+Im[ V_{us}V_{cs}^*V_{ub^\prime}^*V_{cb^\prime}]+Im[V_{ub}V_{cb}V_{ub^\prime}^*V_{cb^\prime}]=0~\text{respectively}.\ee
It may be noted that for other orthogonal condition of $V_{CKM4}$, we get 4 such relations each, yielding a total of 48 relations. These 48 relations lead to 36 different quartets, which now are not all equal. 
After somewhat lengthy calculations, we can check that out of these 36 different quartets, there are only 9 independent ones and  all other quartets are combination of these 9. In the present work, we define these 9 independent rephasing invariants as 
 \begin{align}
 &(J_{4\times 4})_1=Im~[V_{ud}~V_{cs}~V_{us}^*~V_{cd}^*]~,\\
&(J_{4\times 4})_2=Im~[V_{us}~V_{cb}~V_{cs}^*~V_{ub}^*]~,\\
&(J_{4\times 4})_3=Im~[V_{ud}~V_{cb}~V_{cd}^*~V_{ub}^*]~,\\
&(J_{4\times 4})_4=Im~[V_{ud}~V_{ts}~V_{td}^*~V_{us}^*]~,\\
&(J_{4\times 4})_5=Im~[V_{cd}~V_{ts}~V_{td}^*~V_{cs}^*]~,\\
&(J_{4\times 4})_6=Im~[V_{cs}~V_{tb}~V_{cb}^*~V_{ts}^*]~,\\
&(J_{4\times 4})_7=Im~[V_{cd}~V_{tb}~V_{td}^*~V_{cb}^*]~,\\
&(J_{4\times 4})_8=Im~[V_{us}~V_{tb}~V_{ts}^*~V_{ub}^*]~,\\
&(J_{4\times 4})_9=Im~[V_{us}~V_{cb}~V_{cs}^*~V_{ub}^*]~.
 \end{align}
One can also write these 9 invariants in terms of 6 mixing angles and 3 phases of the ${4\times 4}$ quark mixing matrix. In particular, using the  PDG4 representation of this matrix, one gets
\begin{small}
\begin{align*}   (J_{4\times 4})_1=&c_{12} c_{13}^2 c_{14}^2 c_{23} c_{24} s_{12} \left(c_{24} s_{13} s_{23} \sin \left(\sigma _{\text{ub}}\right)-c_{13} s_{14} s_{24} \sin \left(\sigma _{\text{cb}'}-\sigma _{\text{ub}'}\right)\right)\end{align*} 
\begin{align*}(J_{4\times 4})_2= &c_{13} c_{14}^2 c_{24} s_{12} s_{13} \left(c_{12} c_{23} \left(c_{13} c_{24} s_{23} \sin \left(\sigma _{\text{ub}}\right)+s_{13} s_{14} s_{24} \sin \left(\sigma _{\text{cb}'}-\sigma _{\text{ub}'}\right)\right)+s_{12} s_{14} s_{23} s_{24} \sin \left(\sigma _{\text{cb}'}-\sigma _{\text{ub}'}+\sigma _{\text{ub}}\right)\right)\end{align*}
\begin{align*}(J_{4\times 4})_3=&-c_{12} c_{13} c_{14}^2 c_{24} s_{13} \left(c_{12} c_{13}^2 s_{14} s_{23} s_{24} \sin \left(\sigma _{\text{cb}'}-\sigma _{\text{ub}'}+\sigma _{\text{ub}}\right)+s_{13} s_{14} s_{24} \left(c_{12} s_{13} s_{23} \sin \left(\sigma _{\text{cb}'}-\sigma _{\text{ub}'}+\sigma _{\text{ub}}\right)
 \right.\right.  \\&   \left.  \left. 
-c_{23} s_{12} \sin \left(\sigma _{\text{cb}'}-\sigma _{\text{ub}'}\right)\right)-c_{23} c_{24} c_{13} s_{12} s_{23} \sin \left(\sigma _{\text{ub}}\right)\right)\end{align*}
  \begin{align*}(J_{4\times 4})_4=&c_{12} c_{13}^2 c_{14}^2 s_{12} \left(c_{23} \left(c_{34}^2 s_{13} s_{23} \sin \left(\sigma _{\text{ub}}\right)-s_{24} s_{34}^2 \left(c_{13} c_{24} s_{14} \sin \left(\sigma _{\text{cb}'}-\sigma _{\text{ub}'}\right)+s_{13} s_{23} s_{24} \sin \left(\sigma _{\text{ub}}\right)\right)\right)+c_{34} s_{23} s_{34}
  \right. \\&   \left.   
  \left(s_{13} s_{23} s_{24} \sin \left(\sigma _{\text{cb}'}+\sigma _{\text{ub}}\right)-c_{13} c_{24} s_{14} \sin \left(\sigma _{\text{ub}'}\right)\right)+c_{34} c_{23}^2 s_{13} s_{24} s_{34} \left(-\sin \left(\sigma _{\text{ub}}-\sigma _{\text{cb}'}\right)\right)\right)\end{align*}
 \begin{align*}(J_{4\times 4})_5=&c_{23} c_{24} \left(\sin \left(\sigma _{\text{cb}'}\right) c_{24} c_{34} s_{23} \left(c_{23}^2+s_{23}^2\right) s_{24} s_{34} s_{13}^2+c_{13} s_{14} \left(c_{34} \left(\sin \left(\sigma _{\text{ub}}-\sigma _{\text{ub}'}\right) c_{24}^2+\sin \left(\sigma _{\text{ub}}+2 \sigma _{\text{cb}'}-\sigma _{\text{ub}'}\right)
   \right.\right.\right.   \\&    \left.\left.  \left. 
  s_{24}^2\right) s_{34} s_{23}^2+\sin \left(\sigma _{\text{ub}}+\sigma _{\text{cb}'}-\sigma _{\text{ub}'}\right) c_{23} s_{24} \left(c_{34}^2-s_{34}^2\right) s_{23}-\sin \left(\sigma _{\text{ub}}-\sigma _{\text{ub}'}\right) c_{23}^2 c_{34} s_{24}^2 s_{34}\right) s_{13}-\sin \left(\sigma _{\text{cb}'}\right) \right.   \\&   \left. 
   c_{13}^2 c_{24} c_{34} s_{14}^2 s_{23} s_{24} s_{34}\right) c_{12}^2+s_{12} \left(\sin \left(\sigma _{\text{ub}}-\sigma _{\text{cb}'}\right) c_{24}^2 c_{34} s_{13} s_{24} s_{34} c_{23}^4+c_{24} \left(-\sin \left(\sigma _{\text{ub}}\right) c_{13}^2 c_{24} s_{13} s_{23} c_{34}^2 \right.\right.   \\& \left.  \left. 
   +\sin \left(\sigma _{\text{cb}'}-\sigma _{\text{ub}'}\right) c_{13} c_{24}^2 s_{14} s_{24} s_{34}^2+\sin \left(\sigma _{\text{cb}'}-\sigma _{\text{ub}'}\right) c_{13} s_{14} s_{24} \left(s_{24}^2 s_{34}^2-c_{34}^2 s_{13}^2\right)\right) c_{23}^3+c_{34} \left(c_{13}^2 s_{13} s_{24}
   \right.\right.   \\& \left.  \left. 
    \left(2 \cos \left(\sigma _{\text{ub}}-\sigma _{\text{ub}'}\right) \sin \left(\sigma _{\text{cb}'}-\sigma _{\text{ub}'}\right) c_{24}^2-\sin \left(\sigma _{\text{ub}}-\sigma _{\text{cb}'}\right) s_{24}^2\right) s_{14}^2+c_{13} c_{24} s_{23} \left(c_{24}^2 \left(\sin \left(\sigma _{\text{ub}'}\right)
\right.\right. \right.\right.  \\& \left.  \left.    \left. \left.
    -2 \cos \left(\sigma _{\text{ub}}-\sigma _{\text{ub}'}\right) \sin \left(\sigma _{\text{ub}}\right) s_{13}^2\right)-2 \left(\cos \left(\sigma _{\text{ub}}\right) \sin \left(\sigma _{\text{ub}}-\sigma _{\text{ub}'}\right) s_{13}^2+\cos \left(\sigma _{\text{cb}'}\right) \sin \left(\sigma _{\text{cb}'}-\sigma _{\text{ub}'}\right)\right) s_{24}^2\right) s_{14}
 \right.\right.  \\&   \left.  \left.     
    +c_{24}^2 s_{13} \left(\sin \left(\sigma _{\text{ub}}+\sigma _{\text{cb}'}\right) s_{13}^2+\sin \left(\sigma _{\text{ub}}-\sigma _{\text{cb}'}\right)\right) s_{23}^2 s_{24}\right) s_{34} c_{23}^2+\left(\sin \left(\sigma _{\text{ub}}\right) c_{13}^2 s_{13} s_{14}^2 s_{23} s_{34}^2 c_{24}^4-c_{13} s_{14} 
 \right.\right.   \\& \left.  \left.     
    \left(\sin \left(\sigma _{\text{cb}'}-\sigma _{\text{ub}'}\right) c_{13}^2 s_{14}^2+\sin \left(2 \sigma _{\text{ub}}+\sigma _{\text{cb}'}-\sigma _{\text{ub}'}\right) s_{13}^2 s_{23}^2\right) s_{24} s_{34}^2 c_{24}^3-\sin \left(\sigma _{\text{ub}}\right) c_{13}^2 c_{34}^2 s_{13} s_{23}^3 c_{24}^2+c_{13} s_{14} s_{24} 
 \right.\right.  \\&   \left.  \left.     
    \left(c_{34}^2 \left(2 \cos \left(\sigma _{\text{ub}}+\sigma _{\text{cb}'}-\sigma _{\text{ub}'}\right) \sin \left(\sigma _{\text{ub}}\right) s_{13}^2+\sin \left(\sigma _{\text{cb}'}-\sigma _{\text{ub}'}\right)\right) s_{23}^2-\left(\sin \left(\sigma _{\text{cb}'}-\sigma _{\text{ub}'}\right) c_{13}^2 s_{14}^2
 \right.\right.\right.\right.   \\&   \left. \left.\left.  \left.     
    +\sin \left(2 \sigma _{\text{ub}}+\sigma _{\text{cb}'}-\sigma _{\text{ub}'}\right) s_{13}^2 s_{23}^2\right) s_{24}^2 s_{34}^2\right) c_{24}+\sin \left(\sigma _{\text{ub}}\right) c_{13}^2 s_{13} s_{14}^2 s_{23} s_{24}^2 \left(c_{34}^2-s_{24}^2 s_{34}^2\right)\right) c_{23}+c_{34} s_{23}
 \right.   \\&   \left.     
     \left(\sin \left(\sigma _{\text{ub}}+\sigma _{\text{cb}'}\right) s_{13} s_{23} s_{24}-\sin \left(\sigma _{\text{ub}'}\right) c_{13} c_{24} s_{14}\right) \left(c_{24}^2 s_{13}^2 s_{23}^2+2 \cos \left(\sigma _{\text{ub}}+\sigma _{\text{cb}'}-\sigma _{\text{ub}'}\right) c_{13} c_{24} s_{13} s_{14} s_{24} s_{23}
 \right.\right.   \\&   \left.  \left.      
     +c_{13}^2 s_{14}^2 s_{24}^2\right) s_{34}\right) c_{12}-c_{23} c_{24} s_{12}^2 \left(\sin \left(\sigma _{\text{cb}'}\right) c_{24} c_{34} s_{23} \left(c_{23}^2+s_{23}^2\right) s_{24} s_{34} s_{13}^2+c_{13} s_{14} \left(c_{34} \left(\sin \left(\sigma _{\text{ub}}-\sigma _{\text{ub}'}\right) c_{24}^2
 \right.\right.\right.   \\&    \left.\left.  \left.      
     +\sin \left(\sigma _{\text{ub}}+2 \sigma _{\text{cb}'}-\sigma _{\text{ub}'}\right) s_{24}^2\right) s_{34} s_{23}^2+\sin \left(\sigma _{\text{ub}}+\sigma _{\text{cb}'}-\sigma _{\text{ub}'}\right) c_{23} s_{24} \left(c_{34}^2-s_{34}^2\right) s_{23}-\sin \left(\sigma _{\text{ub}}-\sigma _{\text{ub}'}\right)
 \right.\right.   \\& \left.  \left.      
      c_{23}^2 c_{34} s_{24}^2 s_{34}\right) s_{13}-\sin \left(\sigma _{\text{cb}'}\right) c_{13}^2 c_{24} c_{34} s_{14}^2 s_{23} s_{24} s_{34}\right) \end{align*}
\begin{align*}(J_{4\times 4})_6=&
c_{23} c_{24} \left(\sin \left(\sigma _{\text{cb}'}\right) c_{24} c_{34} s_{23} \left(c_{23}^2+s_{23}^2\right) s_{24} s_{34} c_{13}^2+s_{13} s_{14} \left(-c_{34} \left(\sin \left(\sigma _{\text{ub}}-\sigma _{\text{ub}'}\right) c_{24}^2+\sin \left(\sigma _{\text{ub}}+2 \sigma _{\text{cb}'}-\sigma _{\text{ub}'}\right)
 \right.\right.  \right. \\& \left.\left.  \left.   
 s_{24}^2\right) s_{34} s_{23}^2+\sin \left(\sigma _{\text{ub}}+\sigma _{\text{cb}'}-\sigma _{\text{ub}'}\right) c_{23} s_{24} \left(s_{34}^2-c_{34}^2\right) s_{23}+\sin \left(\sigma _{\text{ub}}-\sigma _{\text{ub}'}\right) c_{23}^2 c_{34} s_{24}^2 s_{34}\right) c_{13}-\sin \left(\sigma _{\text{cb}'}\right)
 \right.   \\&   \left.    
  c_{24} c_{34} s_{13}^2 s_{14}^2 s_{23} s_{24} s_{34}\right) c_{12}^2+s_{12} \left(c_{24} s_{14} s_{23} \left(-c_{34} \left(\sin \left(\sigma _{\text{ub}'}\right) c_{24}^2-\sin \left(2 \sigma _{\text{cb}'}-\sigma _{\text{ub}'}\right) s_{24}^2\right) s_{34} c_{23}^2
 \right.\right.   \\& \left.  \left.     
  +\sin \left(\sigma _{\text{cb}'}-\sigma _{\text{ub}'}\right) s_{23} s_{24} \left(s_{34}^2-c_{34}^2\right) c_{23}+\sin \left(\sigma _{\text{ub}'}\right) c_{34} s_{23}^2 s_{24}^2 s_{34}\right) c_{13}^3+s_{13} \left(\sin \left(\sigma _{\text{ub}}\right) c_{24}^2 c_{34}^2 s_{23} c_{23}^3+c_{34} s_{24} 
 \right.\right.   \\& \left.  \left.     
  \left(\left(\sin \left(\sigma _{\text{ub}}+\sigma _{\text{cb}'}\right) s_{23}^2-\sin \left(\sigma _{\text{ub}}+\sigma _{\text{cb}'}-2 \sigma _{\text{ub}'}\right) s_{14}^2\right) c_{24}^2+\sin \left(\sigma _{\text{ub}}-\sigma _{\text{cb}'}\right) s_{14}^2 s_{24}^2\right) s_{34} c_{23}^2+\sin \left(\sigma _{\text{ub}}\right) s_{23}
 \right.\right.   \\& \left.  \left.     
   \left(-s_{14}^2 s_{34}^2 c_{24}^4+c_{34}^2 s_{23}^2 c_{24}^2+s_{14}^2 s_{24}^2 \left(s_{24}^2 s_{34}^2-c_{34}^2\right)\right) c_{23}+c_{34} s_{23}^2 s_{24} \left(c_{24}^2 \left(\sin \left(\sigma _{\text{ub}}+\sigma _{\text{cb}'}\right) s_{23}^2
 \right.\right. \right.\right.  \\& \left. \left. \left.\left.      
   -\sin \left(\sigma _{\text{ub}}+\sigma _{\text{cb}'}-2 \sigma _{\text{ub}'}\right) s_{14}^2\right)-\sin \left(\sigma _{\text{ub}}+\sigma _{\text{cb}'}\right) s_{14}^2 s_{24}^2\right) s_{34}\right) c_{13}^2+c_{24} s_{13}^2 s_{14} \left(\sin \left(\sigma _{\text{cb}'}-\sigma _{\text{ub}'}\right) c_{34}^2 s_{24} c_{23}^3
 \right.\right.   \\& \left.  \left.      
   +c_{34} s_{23} \left(\sin \left(2 \sigma _{\text{ub}}-\sigma _{\text{ub}'}\right) c_{24}^2+\left(2 \cos \left(\sigma _{\text{ub}}\right) \sin \left(\sigma _{\text{ub}}-\sigma _{\text{ub}'}\right)+\sin \left(2 \sigma _{\text{cb}'}-\sigma _{\text{ub}'}\right)\right) s_{24}^2\right) s_{34} c_{23}^2-s_{24}
 \right.\right.   \\& \left.  \left.      
    \left(\sin \left(2 \sigma _{\text{ub}}+\sigma _{\text{cb}'}-\sigma _{\text{ub}'}\right) c_{34}^2 s_{23}^2+\left(\sin \left(\sigma _{\text{cb}'}-\sigma _{\text{ub}'}\right) s_{14}^2-2 \cos \left(\sigma _{\text{ub}}\right) \sin \left(\sigma _{\text{ub}}+\sigma _{\text{cb}'}-\sigma _{\text{ub}'}\right) s_{23}^2\right) s_{34}^2\right) c_{23}
 \right.\right.   \\& \left.  \left.       
    +c_{34} s_{23} \left(\sin \left(\sigma _{\text{ub}'}\right) c_{24}^2 s_{23}^2-\left(\sin \left(\sigma _{\text{ub}'}\right) s_{14}^2+\sin \left(2 \sigma _{\text{ub}}+2 \sigma _{\text{cb}'}-\sigma _{\text{ub}'}\right) s_{23}^2\right) s_{24}^2\right) s_{34}\right) c_{13}+c_{24}^2 c_{34} s_{13}^3 s_{14}^2
 \right.  \\& \left.        
     \left(-\sin \left(\sigma _{\text{ub}}-\sigma _{\text{cb}'}\right) c_{23}^2-\sin \left(\sigma _{\text{ub}}+\sigma _{\text{cb}'}\right) s_{23}^2\right) s_{24} s_{34}\right) c_{12}+c_{24} s_{12}^2 s_{14} \left(-\sin \left(\sigma _{\text{cb}'}\right) c_{23} c_{24} c_{34} s_{14} s_{23} s_{24} s_{34} c_{13}^2
 \right.  \\&   \left.        
     +s_{13} \left(\sin \left(\sigma _{\text{ub}}+\sigma _{\text{cb}'}-\sigma _{\text{ub}'}\right) c_{23}^2 s_{23} s_{24} c_{34}^2+c_{23} \left(\sin \left(\sigma _{\text{ub}}-\sigma _{\text{ub}'}\right) c_{24}^2 s_{23}^2+\left(2 \cos \left(\sigma _{\text{cb}'}\right) \sin \left(\sigma _{\text{ub}}+\sigma _{\text{cb}'}-\sigma _{\text{ub}'}\right)
 \right.\right. \right. \right. \\& \left.  \left.\left.     \left.   
      s_{23}^2-\sin \left(\sigma _{\text{ub}}-\sigma _{\text{ub}'}\right) s_{14}^2\right) s_{24}^2\right) s_{34} c_{34}+\sin \left(\sigma _{\text{ub}}+\sigma _{\text{cb}'}-\sigma _{\text{ub}'}\right) s_{23} \left(s_{23}^2-s_{14}^2\right) s_{24} s_{34}^2\right) c_{13}+\sin \left(\sigma _{\text{cb}'}\right) 
 \right.  \\&   \left.         
      c_{23} c_{24} c_{34} s_{13}^2 s_{14} s_{23} s_{24} s_{34}\right)\end{align*} 
 \begin{align*}(J_{4\times 4})_7=&s_{12} \left(c_{24} s_{14} s_{23} \left(-c_{34} \left(\sin \left(\sigma _{\text{ub}'}\right) c_{24}^2-\sin \left(2 \sigma _{\text{cb}'}-\sigma _{\text{ub}'}\right) s_{24}^2\right) s_{34} c_{23}^2+\sin \left(\sigma _{\text{cb}'}-\sigma _{\text{ub}'}\right) s_{23} s_{24} \left(s_{34}^2-c_{34}^2\right) c_{23}
 \right.\right.  \\&   \left.  \left.  
 +\sin \left(\sigma _{\text{ub}'}\right) c_{34} s_{23}^2 s_{24}^2 s_{34}\right) c_{13}^3+s_{13} \left(\sin \left(\sigma _{\text{ub}}\right) c_{24}^2 c_{34}^2 s_{23} c_{23}^3+c_{34} s_{24} \left(\left(\sin \left(\sigma _{\text{ub}}+\sigma _{\text{cb}'}\right) s_{23}^2
 \right.\right.\right.\right.  \\&   \left.\left.  \left.\left.  
 -\sin \left(\sigma _{\text{ub}}+\sigma _{\text{cb}'}-2 \sigma _{\text{ub}'}\right) s_{14}^2\right) c_{24}^2+\sin \left(\sigma _{\text{ub}}-\sigma _{\text{cb}'}\right) s_{14}^2 s_{24}^2\right) s_{34} c_{23}^2+\sin \left(\sigma _{\text{ub}}\right) s_{23} \left(-s_{14}^2 s_{34}^2 c_{24}^4+c_{34}^2 s_{23}^2 c_{24}^2
 \right.\right. \right. \\&  \left. \left.  \left.  
 +s_{14}^2 s_{24}^2 \left(s_{24}^2 s_{34}^2-c_{34}^2\right)\right) c_{23}+c_{34} s_{23}^2 s_{24} \left(c_{24}^2 \left(\sin \left(\sigma _{\text{ub}}+\sigma _{\text{cb}'}\right) s_{23}^2-\sin \left(\sigma _{\text{ub}}+\sigma _{\text{cb}'}-2 \sigma _{\text{ub}'}\right) s_{14}^2\right)
 \right.\right.\right.  \\& \left.  \left.  \left.  
 -\sin \left(\sigma _{\text{ub}}+\sigma _{\text{cb}'}\right) s_{14}^2 s_{24}^2\right) s_{34}\right) c_{13}^2+c_{24} s_{13}^2 s_{14} \left(\sin \left(\sigma _{\text{cb}'}-\sigma _{\text{ub}'}\right) c_{34}^2 s_{24} c_{23}^3+c_{34} s_{23} \left(\sin \left(2 \sigma _{\text{ub}}-\sigma _{\text{ub}'}\right) c_{24}^2
 \right.\right.\right.  \\& \left.  \left.  \left.  
 +\left(2 \cos \left(\sigma _{\text{ub}}\right) \sin \left(\sigma _{\text{ub}}-\sigma _{\text{ub}'}\right)+\sin \left(2 \sigma _{\text{cb}'}-\sigma _{\text{ub}'}\right)\right) s_{24}^2\right) s_{34} c_{23}^2-s_{24} \left(\sin \left(2 \sigma _{\text{ub}}+\sigma _{\text{cb}'}-\sigma _{\text{ub}'}\right) c_{34}^2 s_{23}^2
 \right.\right.\right.  \\&   \left. \left. \left.  
 +\left(\sin \left(\sigma _{\text{cb}'}-\sigma _{\text{ub}'}\right) s_{14}^2-2 \cos \left(\sigma _{\text{ub}}\right) \sin \left(\sigma _{\text{ub}}+\sigma _{\text{cb}'}-\sigma _{\text{ub}'}\right) s_{23}^2\right) s_{34}^2\right) c_{23}+c_{34} s_{23} \left(\sin \left(\sigma _{\text{ub}'}\right) c_{24}^2 s_{23}^2
 \right.\right.\right.  \\&  \left. \left.  \left.  
 -\left(\sin \left(\sigma _{\text{ub}'}\right) s_{14}^2+\sin \left(2 \sigma _{\text{ub}}+2 \sigma _{\text{cb}'}-\sigma _{\text{ub}'}\right) s_{23}^2\right) s_{24}^2\right) s_{34}\right) c_{13}+c_{24}^2 c_{34} s_{13}^3 s_{14}^2 \left(-\sin \left(\sigma _{\text{ub}}-\sigma _{\text{cb}'}\right) c_{23}^2-
 \right.\right.  \\&   \left.  \left.  
 \sin \left(\sigma _{\text{ub}}+\sigma _{\text{cb}'}\right) s_{23}^2\right) s_{24} s_{34}\right) c_{12}-c_{12}^2 c_{24} s_{14} \left(-\sin \left(\sigma _{\text{cb}'}\right) c_{23} c_{24} c_{34} s_{14} s_{23} s_{24} s_{34} c_{13}^2+s_{13} \left(\sin \left(\sigma _{\text{ub}}+\sigma _{\text{cb}'}-\sigma _{\text{ub}'}\right)
 \right.\right.  \\&   \left.  \left.  
  c_{23}^2 s_{23} s_{24} c_{34}^2+c_{23} \left(\sin \left(\sigma _{\text{ub}}-\sigma _{\text{ub}'}\right) c_{24}^2 s_{23}^2+\left(2 \cos \left(\sigma _{\text{cb}'}\right) \sin \left(\sigma _{\text{ub}}+\sigma _{\text{cb}'}-\sigma _{\text{ub}'}\right) s_{23}^2-\sin \left(\sigma _{\text{ub}}-\sigma _{\text{ub}'}\right) s_{14}^2\right)
 \right.\right. \right. \\&   \left.  \left.   \left.
   s_{24}^2\right) s_{34} c_{34}+\sin \left(\sigma _{\text{ub}}+\sigma _{\text{cb}'}-\sigma _{\text{ub}'}\right) s_{23} \left(s_{23}^2-s_{14}^2\right) s_{24} s_{34}^2\right) c_{13}+\sin \left(\sigma _{\text{cb}'}\right) c_{23} c_{24} c_{34} s_{13}^2 s_{14} s_{23} s_{24} s_{34}\right)-c_{23} c_{24}
 \\&     
    s_{12}^2 \left(\sin \left(\sigma _{\text{cb}'}\right) c_{24} c_{34} s_{23} \left(c_{23}^2+s_{23}^2\right) s_{24} s_{34} c_{13}^2+s_{13} s_{14} \left(-c_{34} \left(\sin \left(\sigma _{\text{ub}}-\sigma _{\text{ub}'}\right) c_{24}^2+\sin \left(\sigma _{\text{ub}}+2 \sigma _{\text{cb}'}-\sigma _{\text{ub}'}\right)
 \right.\right. \right. \\&  \left. \left.  \left.     
     s_{24}^2\right) s_{34} s_{23}^2+\sin \left(\sigma _{\text{ub}}+\sigma _{\text{cb}'}-\sigma _{\text{ub}'}\right) c_{23} s_{24} \left(s_{34}^2-c_{34}^2\right) s_{23}+\sin \left(\sigma _{\text{ub}}-\sigma _{\text{ub}'}\right) c_{23}^2 c_{34} s_{24}^2 s_{34}\right) c_{13}-\sin \left(\sigma _{\text{cb}'}\right)
 \right.  \\&   \left.      
      c_{24} c_{34} s_{13}^2 s_{14}^2 s_{23} s_{24} s_{34}\right)\end{align*}
\begin{align*}(J_{4\times 4})_8=&c_{13} c_{14}^2 s_{12} s_{13} \left(c_{24} s_{12} s_{14} s_{34} \left(c_{23} c_{34} \sin \left(\sigma _{\text{ub}}-\sigma _{\text{ub}'}\right)+s_{23} s_{24} s_{34} \sin \left(\sigma _{\text{cb}'}-\sigma _{\text{ub}'}+\sigma _{\text{ub}}\right)\right)+c_{12} \left(c_{24} s_{13} s_{14} s_{34} 
 \right.\right. \\&  \left. \left.   
\left(c_{23} s_{24} s_{34} \sin \left(\sigma _{\text{cb}'}-\sigma _{\text{ub}'}\right)+c_{34} s_{23} \sin \left(\sigma _{\text{ub}'}\right)\right)+c_{13} \left(c_{34} c_{23}^2 s_{24} s_{34} \left(-\sin \left(\sigma _{\text{ub}}-\sigma _{\text{cb}'}\right)\right)+c_{34} s_{23}^2 s_{24} s_{34}
 \right. \right.\right.\\& \left. \left. \left.   
 \sin \left(\sigma _{\text{cb}'}+\sigma _{\text{ub}}\right)+c_{23} s_{23} \left(c_{34}^2-s_{24}^2 s_{34}^2\right) \sin \left(\sigma _{\text{ub}}\right)\right)\right)\right) \end{align*}
\begin{align*}(J_{4\times 4})_9=&c_{12} c_{13} c_{14}^2 s_{13} \left(c_{12} c_{24} c_{13}^2 s_{14} s_{34} \left(c_{23} c_{34} \sin \left(\sigma _{\text{ub}}-\sigma _{\text{ub}'}\right)+s_{23} s_{24} s_{34} \sin \left(\sigma _{\text{cb}'}-\sigma _{\text{ub}'}+\sigma _{\text{ub}}\right)\right)+c_{24} s_{13} s_{14} s_{34}
\right. \\& \left.
 \left(c_{12} s_{13} \left(c_{23} c_{34} \sin \left(\sigma _{\text{ub}}-\sigma _{\text{ub}'}\right)+s_{23} s_{24} s_{34} \sin \left(\sigma _{\text{cb}'}-\sigma _{\text{ub}'}+\sigma _{\text{ub}}\right)\right)-s_{12} \left(c_{23} s_{24} s_{34} \sin \left(\sigma _{\text{cb}'}-\sigma _{\text{ub}'}\right)+c_{34} s_{23}
\right. \right.\right.\\& \left. \left. \left.   
  \sin \left(\sigma _{\text{ub}'}\right)\right)\right)-c_{13} s_{12} \left(c_{34} c_{23}^2 s_{24} s_{34} \left(-\sin \left(\sigma _{\text{ub}}-\sigma _{\text{cb}'}\right)\right)+c_{34} s_{23}^2 s_{24} s_{34} \sin \left(\sigma _{\text{cb}'}+\sigma _{\text{ub}}\right)+c_{23} s_{23} \left(c_{34}^2-s_{24}^2 s_{34}^2\right)
\right. \right.\\&  \left. \left.    
   \sin \left(\sigma _{\text{ub}}\right)\right)\right)\end{align*}
 
\end{small}

As a next step, we have made an attempt to numerically evaluate the above mentioned 9 independent $J_{4\times 4}$ rephasing invariants. To this end, one needs to consider the 6 mixing angles as well as the 3 phases as inputs. As mentioned earlier, the PDG4 representation of ${4\times 4}$ quark mixing matrix can be reduced to the PDG representation corresponding to the 3 generation case. Therefore, for this representation, considering, since
$ s_{12} \cong V_{us}, ~~ s_{13} \cong V_{ub},~~s_{23} \cong V_{cb}$, using the experimentally well determined values of these mixing elements \cite{pdg24}, we get
\be  \theta_{12}= 0.2263\pm 0.0009,~~\theta_{13}=0.0038\pm 0.0002,~~\theta_{23}=0.0410\pm 0.0010.\label{s12s23s13}\ee
Further, the CP violating phase of this representation, usually called $\delta$, referred to as $\sigma_{ub}$ here for the PDG4 representation, is known to be \cite{pdg24}
\be \sigma_{ub} = 1.24 \pm 0.23. \label{delta}\ee

\begin{table}[h!]
\renewcommand{\arraystretch}{1.5}
\centering
\caption{Numerical constraints on $4^{th}$ row and $4^{th}$ column  elements}
\vspace{0.5cm}
\begin{tabular}{|c|c|c|}
\hline
S.No. & Parameter &  Value\\ \hline
1& $V_{ub^\prime}$ & $0.017 \pm 0.014$\\ \hline
2& $V_{cb^\prime}$ & $(8.4\pm6.2)\times 10^{-3}$\\ \hline
3& $V_{tb^\prime}$ & $0.07 \pm 0.08$\\ \hline
4& $V_{t^\prime d}$ & $0.01 \pm 0.01$\\ \hline
5& $V_{t^\prime s}$ & $0.01 \pm 0.01$\\ \hline
6& $V_{t^\prime b}$ & $0.07 \pm 0.08$\\ \hline
7& $V_{t^\prime b^\prime}$ & $0.998 \pm 0.006$\\ \hline
\end{tabular} 
\label{digheinputs} 
\end{table}

In order to determine the additional mixing angles for the fourth generation case, since we have constructed the PDG4 representation, mentioned earlier by taking the product of the 6 rotation matrices in a particular sequence, we can consider $s_{14} \cong V_{ub^\prime},~~s_{24} \cong V_{cb^\prime},~~s_{34} \cong V_{tb^\prime}$. Numerical constraints for the fourth generation mixing matrix elements as well as the two additional phases $\sigma_{ub^{\prime}}$ and $\sigma_{cb^{\prime}}$ have been provided in Ref. \cite{dighe11}. In particular, performing a fit to the direct unitarity based measurements of some of the elements alongwith measurements of several flavor-
changing observables in the K and B systems that have small hadronic uncertainties and also considering the constraints from the vertex corrections to $Z\rightarrow b\bar{b}$, constraints on the $4^{th}$ row and $4^{th}$ column  elements have been provided. These are mentioned in Table \ref{digheinputs}. From these, one obtains
\begin{align}  \theta_{14}= 0.0170\pm 0.0140,~~\theta_{24}=0.008\pm 0.00624,~~\theta_{34}=0.0700\pm 0.0802.\label{news12s23s13}\end{align}
Further, the two additional phases have been given as \cite{dighe11}
\be \sigma_{ub^\prime} = 1.21 \pm 1.59, ~~~~~~~ \sigma_{cb^\prime} = 1.10 \pm 1.64. \label{newdeltas} \ee

\begin{table}[h!]
\renewcommand{\arraystretch}{1.5}
\centering
\caption{Numerical values of the 9 independent $J_{4\times 4}$ rephasing invariants}
\vspace{0.5cm}
\begin{tabular}{|c|c|c|}
\hline
 $J_{4\times 4}$& Numerical value
  \\ \hline
$(J_{4\times 4})_1$& $(3.687\pm 0.573)\times 10^{-5}$ \\ \hline
$(J_{4\times 4})_2$& $(3.341\pm 0.427)\times 10^{-5}$ \\ \hline
$(J_{4\times 4})_3$& $(3.338\pm 0.427)\times 10^{-5}$ \\ \hline
$(J_{4\times 4})_4$& $(2.312\pm 1.614)\times 10^{-5}$ \\ \hline
$(J_{4\times 4})_5$& $(2.314\pm 1.607)\times 10^{-5}$ \\ \hline
$(J_{4\times 4})_6$& $(4.350\pm 2.754)\times 10^{-5}$ \\ \hline
$(J_{4\times 4})_7$& $(2.209\pm 1.713)\times 10^{-5}$ \\ \hline
$(J_{4\times 4})_8$& $(3.318\pm 0.433)\times 10^{-5}$ \\ \hline
$(J_{4\times 4})_9$& $(3.304\pm0.794)\times 10^{-5}$ \\ \hline
\end{tabular} 
\label{diff j} 
\end{table}

After having obtained the values of the 6 mixing angles and the 3 phases, as mentioned in equations (\ref{s12s23s13}), (\ref{delta}), (\ref{news12s23s13}) and (\ref{newdeltas}), we have evaluated the numerical values of the 9 independent $J_{4\times 4}$ rephasing invariants, presented in Table \ref{diff j}. A look at this table reveals that the numerical values of these 9 invariants vary considerably, i.e., these vary from $(2.209\pm 1.713)\times 10^{-5}$ to $(4.350\pm 2.754)\times 10^{-5}$. Interestingly, the rephasing invariant parameter in the case of 3 generations being $(3.12^{+0.13}_{-0.12})\times 10^{-5}$ is included in the numerical ranges of these 9 rephasing invariants parameters found here. These values may provide valuable clues regarding the 4$^{th}$ generation as the inputs become more refined. 

\section{Summary and Conclusions}
To summarize, as a first step, in the case of fourth generation, we have constructed the PDG4 representation of the quark mixing matrix which can be easily reduced to the PDG representation of the 3$\times$3 quark mixing matrix. Using unitarity constraints as well as the hierarchy among the elements of the CKM matrix, we have found the hierarchy among the 4$^{th}$ row and 4$^{th}$ column elements of the $4\times 4$ mixing matrix. Further, we find that unlike the case of 3 generations wherein there is a unique Jarlskog's rephasing invariant parameter J, in the fourth generation case, there are 9 independent $J_{4\times 4}$ rephasing invariants. Using the PDG4 representation of the $4\times 4$ mixing matrix, we have found explicit expressions for the 9 independent $J_{4\times 4}$. Furthermore, using phenomenological estimates of the 4$^{th}$ row and 4$^{th}$ column elements, we have numerically evaluated these 9 parameters.

\section*{Acknowledgements}
The authors would like to thank the Chairperson, Department of Physics, Panjab University, Chandigarh, for providing the facilities to work.

\end{document}